\documentclass[twocolumn,amssymb,aps,pre,showpacs,nofootinbib,10pt,superscriptaddress]{revtex4-1}
\usepackage{graphicx}
\usepackage{hyperref}
\usepackage{dcolumn}
\usepackage{bm}
\usepackage{latexsym}
\usepackage{amsmath}
\usepackage{amssymb}
\usepackage{commath}
\usepackage[caption=false]{subfig}
\usepackage{color}  
\usepackage[normalem]{ulem}  

\newcommand{\Nc}{N_{\rm c}}
\newcommand{\Ng}{N_{\rm g}}
\newcommand{\Fth}{F_{\rm th}}
\newcommand{\Fmax}{F_{\rm max}}
\newcommand{\fth}{f_{\rm th}}

\newcommand{\bp}{\bar{p}}
\newcommand{\NL}{N_{\rm L}}  
\newcommand{\NE}{N_{\rm E}}  

\newcommand{\blue}[1]{\textcolor{blue}{#1}}

\newcommand{\vspacebeforecaption}{\vspace{-.5cm}}
\newcommand{\vspaceaftercaption}{\vspace{-.2cm}}

\begin{document}
\bibliographystyle{apsrev4-1}
\title{Cascading Failures in AC Electricity Grids}

\author{Martin Rohden}
\email{m.rohden@jacobs-university.de}
\affiliation{%
    Jacobs University, Department of Physics and Earth Sciences,
    Campus Ring 1, 28759 Bremen, Germany.
}

\author{Daniel Jung}
\email{d.jung@jacobs-university.de}
\affiliation{%
    Jacobs University, Department of Physics and Earth Sciences,
    Campus Ring 1, 28759 Bremen, Germany.
}

\author{Samyak Tamrakar}
\email{samyak.r.tamrakar@gmail.com}
\affiliation{%
    Jacobs University, Department of Physics and Earth Sciences,
    Campus Ring 1, 28759 Bremen, Germany.
}
\affiliation{
    Physics Department, Carl von Ossietzky Universit\"at Oldenburg,
    Ammerl\"ander Heerstra\ss e 114-118, 26129 Oldenburg.
}

\author{Stefan Kettemann}
\email{s.kettemann@jacobs-university.de}
\affiliation{%
    Jacobs University, Department of Physics and Earth Sciences,
    Campus Ring 1, 28759 Bremen, Germany.
}
\affiliation{%
    Pohang University of Science and Technology, Division of Advanced Materials
    Science, San 31, Hyoja-dong, Nam-gu, Pohang 790-784, South Korea.
}

\date{\today}

\begin{abstract}

Sudden failure of a single transmission element in a power grid can induce a
domino effect of cascading failures, which can lead to the isolation of a large
number of consumers or even to the failure of the entire grid. Here we present
results of the simulation of cascading failures in power grids, using an
alternating current (AC) model. We first apply this model to a regular
square grid topology. For a random placement of consumers and
generators on the grid, the probability to find more than a
certain number of unsupplied consumers decays as a power law and obeys a
scaling law with respect to system size. Varying the transmitted power threshold
above which a transmission line fails does not seem to change the power law exponent
$q \approx 1.6$. Furthermore, we study the influence of the placement of generators and
consumers on the number of affected consumers and demonstrate that large clusters
of generators and consumers are especially vulnerable to cascading failures. As a
real-world topology we consider the German high-voltage transmission grid. Applying
the dynamic AC model and considering a random placement of consumers, we find
that the probability to disconnect more than a certain number of consumers
depends strongly on the threshold. For large thresholds the decay is clearly
exponential, while for small ones the decay is slow, indicating a power
law decay.

\end{abstract}


\maketitle


A reliable supply of electric power is of fundamental importance for the
technical infrastructure of modern societies. In fact, the reliability of
electric power grids has increased continuously in the last decades
\cite{bundesnetzagentur}. However, large-scale power outages still occur and
can affect millions of customers which may result in catastrophic events. It is
therefore important to understand which topological properties of power grids
and which placements of generators and consumers on the grid are able to
diminish the risk of large-scale outages.

Large-scale outages can often be traced back to the failure of a single
transmission element \cite{outage1,outage2,outage3,outage4}. The initial
failure causes secondary failures, which can eventually lead to a whole cascade
of failures. Cascading failures have been analyzed in various studies with
different models and from different viewpoints
\cite{motter,AB1,AB2,SS,WW,RL,YK,BC,ID,BP,scala,Simonsen,Plietzsch}. Most of these previous
studies analyze the influence of network topology on the cascade of failures
using simplified topological flow models such as the \emph{messenger model}
introduced by Motter and Lai \cite{motter}, which is used to study cascading
failures in Refs.~\cite{witthaut13,witthaut15}.

In this work we base the analysis of cascading failures on the \emph{AC
power flow equations} \cite{hill,filatrella,timme,rohdenchaos}. This allows us
to study the influence of the physical properties of the grid, such as the
placement of generators and consumers and the power capacitance of the
transmission lines, on the probability and the extent of cascading failures. To
this end, we consider a random placement of generators and consumers on a
regular 2D grid graph (square grid) as well as on a model topology for the German high-voltage
transmission grid ($380\,\text{kV}$) \cite{scigrid}.

An evaluation of the statistics of power failures which occurred in real power
grids in the last decades shows that the probability of ending up with more than
a certain number of unsupplied consumers often decays like a \emph{power law} with the
number of unsupplied consumers $\Nc$ \cite{ID,BP}. Such a slow decay indicates a
significant probability that a large number of consumers become unsupplied. It is
therefore of practical relevance to understand which properties of the grid are responsible for this behavior. In
Ref.~\cite{ID}, a simple model has been suggested to simulate cascading
failures that assumes that the load $F_{ij}$ of a failing transmission line is
redistributed in equal parts among the remaining transmission lines. This model
can be solved analytically and yields a \emph{power law} distribution when the
initial average power flow through the transmission lines prior to the failure
$F = \langle F_{ij} \rangle$ reaches a certain critical ratio of the threshold
power $\Fth$. Below that value, an exponentially fast decay is found \cite{ID}.

While most previous studies assume that the power law dependence is only
related to the network topology, in particular to scale-free topologies
\cite{Cuadra2015,Goh2003}, the main goal of this work is to analyze which influence
different placements of consumers and generators have on the probability to
find $\Nc$ unsupplied consumers. We start by finding the stationary power
flow in the fully connected grid. Then we initiate a power line failure by
removing one transmission line by hand and find the new stationary power flow.
The resulting redistribution of the power flow may trigger further line failures where
the transmitted power exceeds a threshold $\Fth$, which we set to be a certain
ratio of the transmission line power capacitance. This chain of outages
continues until we cannot find a stable solution for the stationary power flow anymore. We then record
the number of consumers $\Nc$ that cannot be supplied anymore by the available generators.
Note that consumers which are connected within clusters might also not be
supplied if the number of generators does not at least equal the number of
consumers in the specific cluster. This process is repeated by subsequently
initiating the power outage with every transmission line of the grid, removing
that line and observing the resulting cascade. Thereby, we obtain the
probability distribution of the \emph{blackout size} $\Nc$ in dependence of the
placement of consumers and generators and the threshold $\Fth$.

We gain further insights by analyzing the share of links that can induce a
cascade of failures depending on the existence of various fixed cluster sizes. We
demonstrate that the existence of large clusters of generators and consumers
makes the grid particularly vulnerable to cascading failures, since the
likelihood for a whole cluster to break down at once appears to increase with
increasing cluster size. We thereby show that the size of the power outage
depends essentially on the placement of consumers and generators. Finally,
we study a real-world topology, a model for the German high-voltage
transmission grid \cite{scigrid}. In this irregular grid structure, we find a
decay of $\bp(\Nc)$ for large $\Nc$ which depends strongly on the threshold
$\Fth$. For large $\Fth$ the decay is clearly exponential. For small $\Fth$ the
decay is slow and may indicate a power law decay. Thus, there might be a
critical value of the threshold $\Fth$ in the German grid below which the
cumulative probability density becomes critical and decays with a power law.


\textit{Power Grid Model.---} We approximate the power grid as a network of $N$
rotating synchronous machines, representing generators and motors. Each machine
$k \in \{1,\ldots,N\}$ is characterized by the net mechanical power $P_k^{\rm
mech}$, which is positive for a generator and negative for a consumer. The
state of machine $k$ is given by the angular frequency and the rotor angle
(power angle) $\theta_k(t)$ which is measured relatively to a reference machine
rotating at the nominal grid angular frequency $\omega_0 = 2\pi \times
50\,\text{Hz}$. Correspondingly, $\omega_k(t) = \dif\theta_k(t) / \dif t =
\dot\theta_k$ gives the angular frequency deviation from the reference
frequency $\omega_0$. The dynamics of the rotors are governed by the
\emph{swing equation} \cite{hill,timme,filatrella},
\begin{equation}
    I_k \od[2]{\theta_k}{t} + D_k \od{\theta_k}{t}
    = P_k^{\rm mech} - P_k^{\rm el} \quad\rm,
    \label{eq:swing}
\end{equation}
where $P_k^{\rm el}$ is the net electrical power transmitted from adjacent
rotating machines through the transmission lines. $I_k$ is the moment of
inertia of the rotor times $\omega_0$ and $D_k$ measures the damping, which is
mainly due to damper windings \cite{machowski}.

For simplicity we neglect ohmic losses of transmission lines which can be
considered small in high voltage levels \cite{heuck}. Thus, the line admittance
is purely inductive, $Y_{k\ell} = 1/(i \omega L_{k\ell})$, where $L_{k\ell}$ is
the inductance of the line $(k,\ell)$. Then, the magnitude of the voltage is
constant throughout the grid, $|U_k| = U_0 \forall k \in \{1,\ldots,N\}$. For a
common two-pole synchronous machine, the phase of the voltage equals the
mechanical phase of the rotor. The expression for the active electric power
then simplifies to
\begin{equation}
    P_k^{\rm el} = \sum_{\ell=1}^{N} \frac{U_0^2}{\omega L_{k \ell}}
    \sin(\theta_k - \theta_\ell) \quad\rm.
\end{equation}
Substituting this result into the swing equation \eqref{eq:swing} yields the
equations of motion,
\begin{equation}
    I_k \od[2]{\theta_k}{t} + D_k \od{\theta_k}{t}
    = P_k^{\rm mech} - \sum_{\ell=1}^{N} \frac{U_0^2}{\omega L_{k\ell}}
    \sin(\theta_k - \theta_\ell) \quad\rm.
    \label{eq:eom-theta}
\end{equation}
Using the abbreviations
\begin{eqnarray}
    P_k &=& \frac{P_k^{\rm mech} - D_k \omega_0}{I_k} \quad\text{,}\quad
    \alpha_k = \frac{D_k}{I_k} \quad\text{,}\quad \nonumber \\
    K_{k\ell} &=& \frac{U_0^2}{I_k \omega L_{k\ell}} \quad\rm, \nonumber
    \label{eq:abbrev}
\end{eqnarray}
the oscillator model reads
\begin{equation}
    \od[2]{\theta_k}{t} = P_k - \alpha_k \od{\theta_k}{t}
    + \sum_{\ell=1}^{N} K_{k\ell}\, \sin(\theta_\ell - \theta_k) \quad\rm.
    \label{eq:osc-model}
\end{equation}
These equations of motion are widely used to model power grids in power
engineering, where it goes by the name \emph{synchronous motor model}
\cite{nishikawa}. Notably, this model is similar to the \textit{Kuramoto
model}, which is studied extensively in statistical physics
\cite{Kura75,Stro00,Aceb05}. In the Kuramoto model, the inertia term is absent,
so it can be seen as the over-damped limiting case.

In order to find the stationary solutions of the oscillator model \eqref{eq:osc-model}, it is
equivalent to solve the static flow equations,
\begin{equation}
    \label{eq:stat}
    0 = P_k + \sum_{\ell=1}^{N} K_{k\ell}\, \sin(\theta_\ell - \theta_k)
    \quad\rm.
\end{equation}
This can
be done by a standard root-finding algorithm \cite{root}.
Note that regular square grids consisting of loops with four edges, which are
studied in the following, may have multiple
stationary (stable and unstable) solutions \cite{Delabays2016}.
We ensure the stability of a stationary state by checking that the Jacobian
matrix has no positive eigenvalues.

\begin{figure}[tb]
    \subfloat[]{
        \label{fig:sg1}
        \includegraphics[width=.45\columnwidth]{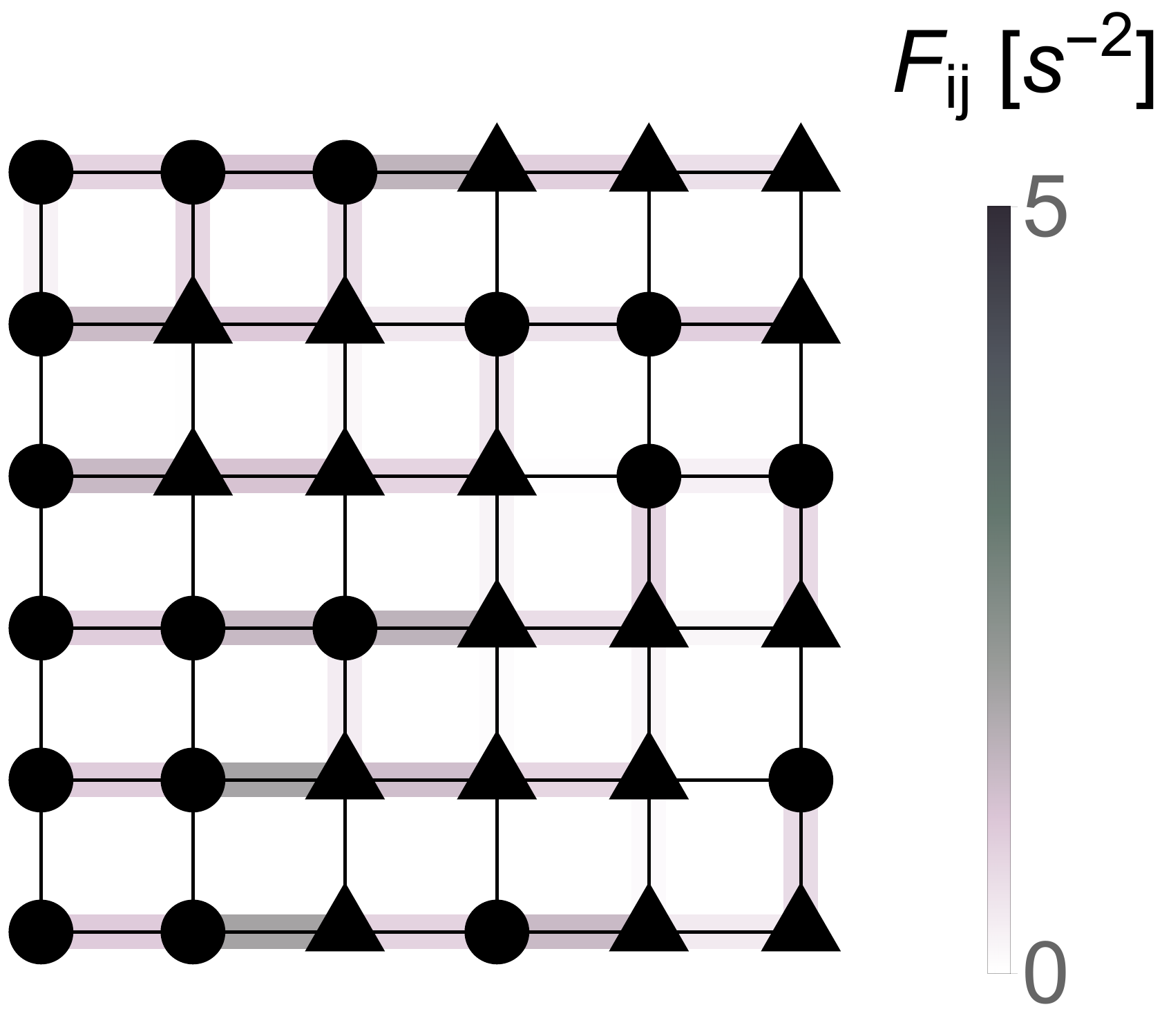}
    }
    \subfloat[]{
        \label{fig:sg2}
        \includegraphics[width=.45\columnwidth]{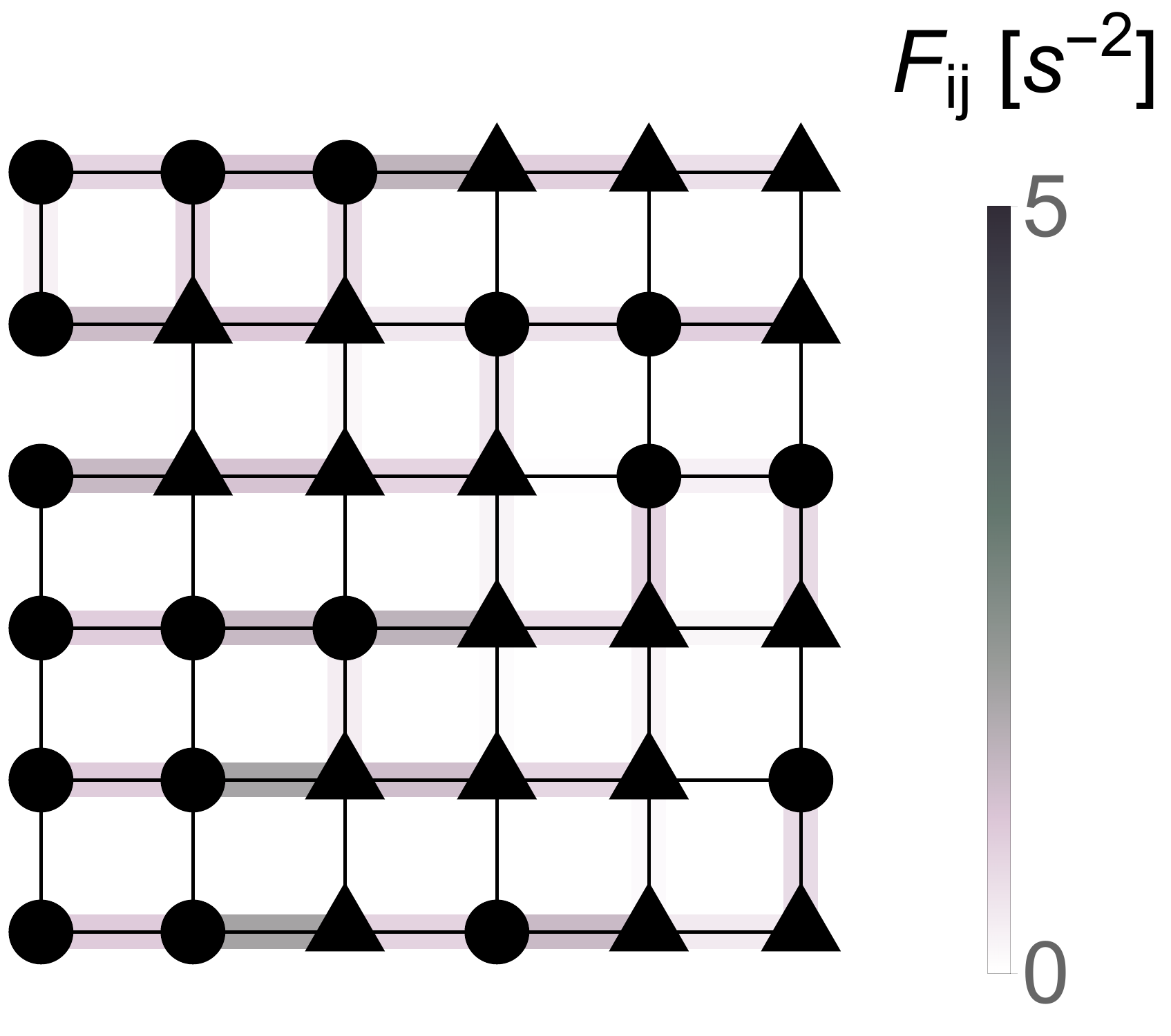}
    }\\
    \subfloat[]{
        \label{fig:sg3}
        \includegraphics[width=.45\columnwidth]{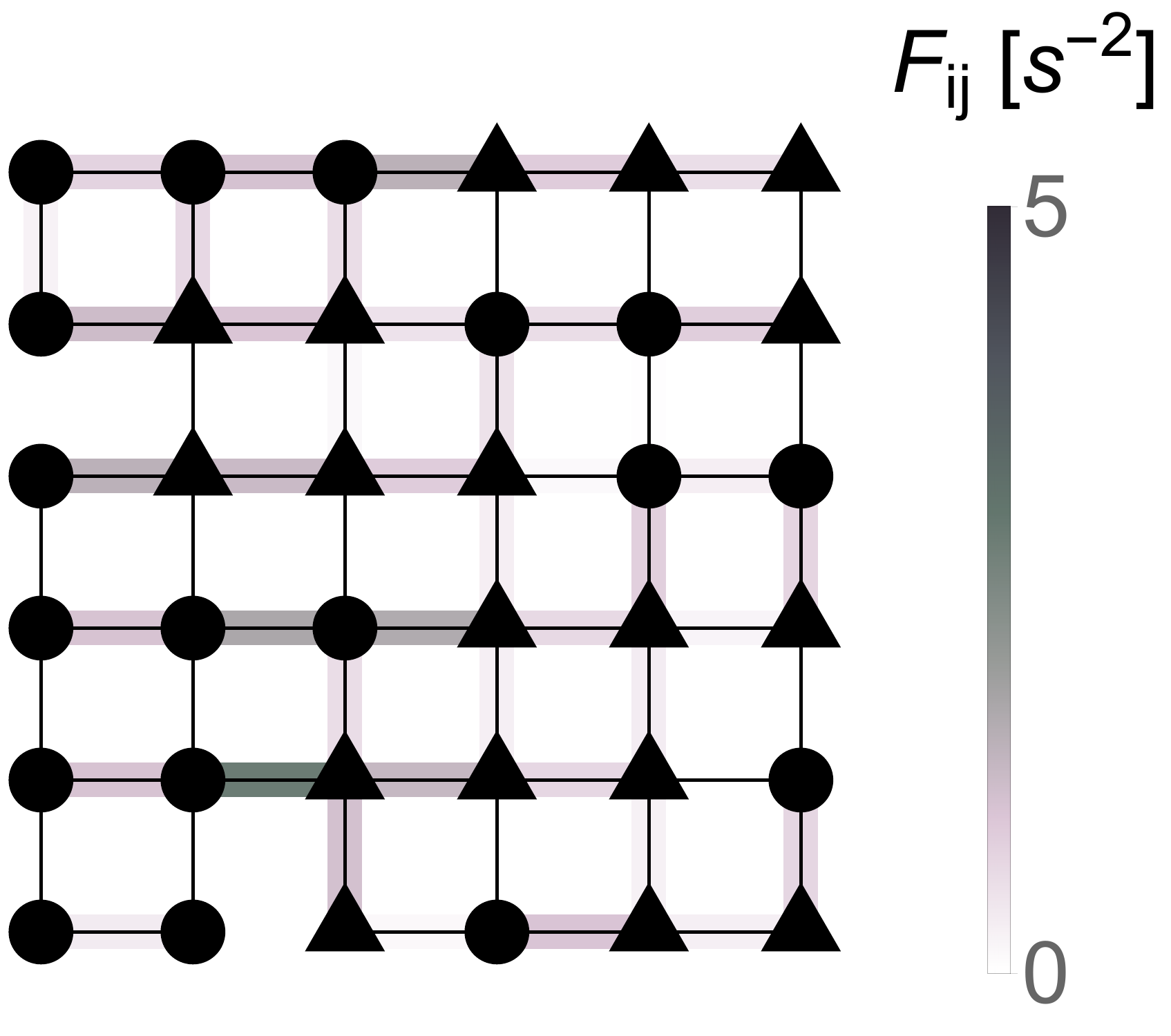}
    }
    \subfloat[]{
        \label{fig:sg4}
        \includegraphics[width=.45\columnwidth]{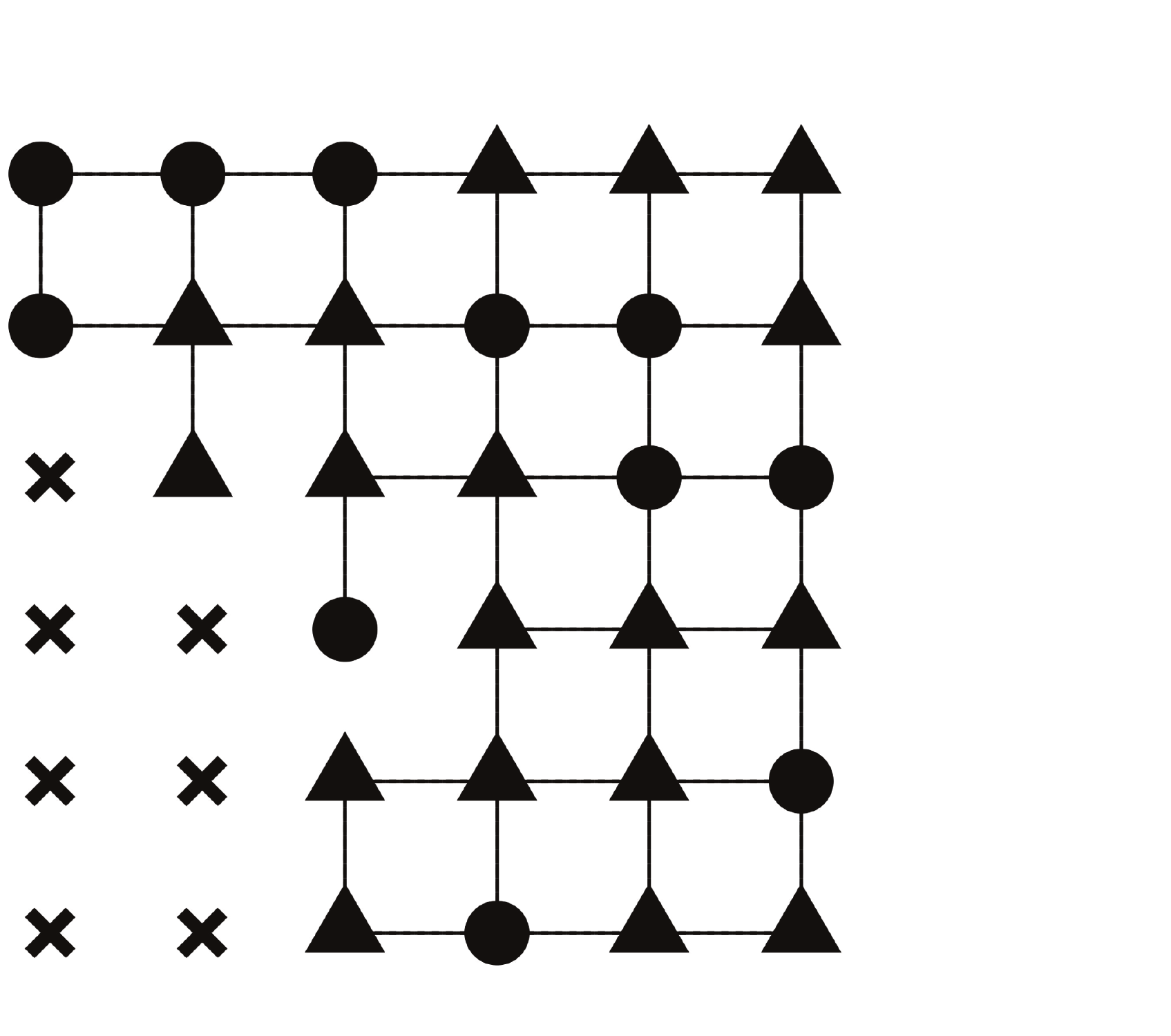}
    }
	\caption{
        (Color online) Example for a cascading failure in a $6 \times 6$ square
        grid with open boundary conditions. Triangles denote generators, disks
        denote consumers. Intact transmission lines are indicated by a thin
        dashed line. The stable state power flow is plotted in color scale
        ranging from $0$ to $5\,{\rm s}^{-2}$. (a) Power flow in the fully
        connected grid. (b) Power flow after removing one line. (c) Power
        flow after removing all lines with a power flow larger than the
        power threshold $\Fth$ after initial removal of a transmission line. (d) After several more steps
        (not shown) some nodes become isolated (indicated by crosses), so the
        simulation is stopped.
    }
    \vspaceaftercaption
	\label{fig:cascade}
\end{figure}

\textit{Cascading Failure Algorithm.---} In the following we describe the
cascading failure algorithm used in this study. We determine the stable state
with power flows $F_{ij}$. Note that we skip realizations for which no stable
state exists or the initial maximal power flow $\max(F_{ij})$ is already larger
than the threshold power flow $\Fth$. Thereby it is ensured that the initial
power flow through all lines is stable. Next, we remove one of the transmission
lines of the network in order to induce a cascade of failures. It is again ensured that the network
reaches a new stable state with a new power flow distribution $F'_{ij}$. All
transmission lines for which the transferred power $F'_{ij}$ exceeds $\Fth$ are
removed from the grid, and the power flows are recalculated. This process is
repeated until no transferred power exceeds $\Fth$ or until the grid splits
into different subgrids. In the latter case we record the number of affected
consumers, the \emph{blackout size} $\Nc$, which is the number of consumers
that cannot be supplied by generators anymore. The whole process is repeated
for each transmission line of the original grid being initially removed, and
also for other placements of generators and consumers $P_k$. Fig.~\ref{fig:cascade}
illustrates the cascading failure algorithm for the example of a $6 \times 6$
square grid. Panel (a) illustrates the initial stable state before the initial
line removed, panel (b) the stable state after the removal of one transmission
line (upper left side of the grid), panel (c) the second step of the cascade of
failures and panel (d) the final step with seven disconnected consumers.

\textit{Statistical Analysis.---} We initiate a cascade of failures by manually
removing each of the $\NL$ transmission lines of the original grid separately.
For a $L \times L$ square grid graph with open boundary conditions, there exist
$\NL = 2L(L-1)$ links, so we perform the cascading failure algorithm $\NL$
times. We repeat this for $R=1000$ realizations of random placements of
generators and consumers $P_k$. For each realization $r$, we obtain the
histogram $E_r(\Nc)$ which counts the number of events that $\Nc$ consumers
are unsupplied. From this, we obtain the normalized \emph{probability distribution function} (pdf)
\begin{equation}
    e_r(\Nc) = \frac{E_r(\Nc)}{\NL} \quad\rm,
    \label{eq:er}
\end{equation}
the share of initially removed transmission lines for which the cascade
resulted in $\Nc$ isolated consumers.

Then, we compute the \emph{complementary cumulative distribution function}
(ccdf) $p_r(\Nc)$ (in short \emph{cumulative probability} in the following),
yielding the probability that the number of unsupplied consumers (the
\emph{blackout size}) is larger than $\Nc$,
\begin{equation}
    p_r(N_{\rm c}) = \sum\limits_{\Nc'=\Nc+1}^{\infty} e_r(\Nc')
    \quad\rm.
    \label{eq:pr}
\end{equation}
Finally, we obtain the ensemble average $\bp(\Nc)$ over $R=1000$ realizations,
\begin{equation}
    \bp(\Nc) = \frac{1}{R} \sum\limits_{r=1}^{R} p_r(\Nc) \quad\rm.
    \label{eq:pm}
\end{equation}

\textit{Regular square grid topology with random placement of consumers.---} In
the following we present the results for the statistics of the number of unsupplied
consumers for a $L \times L$ square grid with open boundary conditions for different linear
system sizes $L$. We consider a simple regular square grid topology with open
boundary conditions this rather simple model to systematically study the influence of different consumer
placements. Half of the nodes serve as consumers and the other half as
generators. Each node $k$ generates the net power $P_k = \pm P$ (positive for
generators, negative for consumers), with $P = 1\,{\rm s}^{-2}$. The power
capacity of all transmission lines is set to $K_{ij} = K = 5\,{\rm s}^{-2}$.
All machines have the same damping parameter $\alpha_k = \alpha = 1\,{\rm
s}^{-1}$.

\begin{figure}[tb]
    \includegraphics[width=\columnwidth]{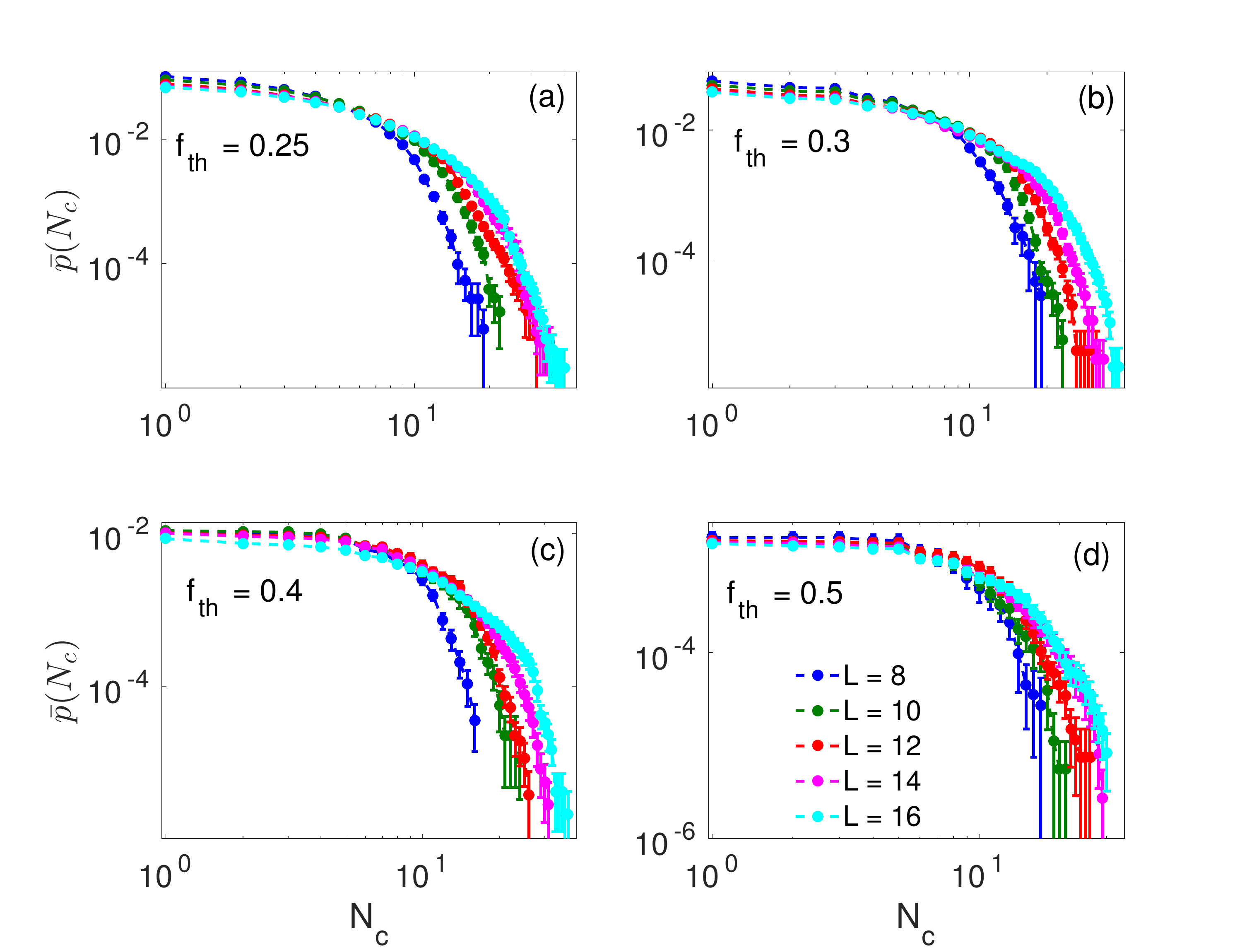}
    \caption{%
        (Color online) Average cumulative probability $\bp(\Nc)$
        (cf.~Eq.~\eqref{eq:pm}) for square grids of different linear system
        size $L$ and threshold $\fth=\Fth/K$: (a) $\fth = 0.25$, (b) $\fth =
        0.3$, (c) $\fth = 0.4$, (d) $\fth = 0.5$. Error bars show the standard error
        (cf.~Eq.~\eqref{eq:pm}).
    }
    \label{fig:pm}
\end{figure}

In order to precisely control the amount of randomness in the system, we use
the following procedure to generate a random array $P_k$
\cite{Labavic2014,Jung2016}: We start from a periodic arrangement of generators
and consumers \cite{antiferro} and divide the graph into two subgraphs, one
carrying all $N/2$ generators and the other all $N/2$ consumers. Then, $p$
different nodes are chosen randomly from each subgraph, forming $p$
generator-consumer pairs. Finally, each of these generator-consumer pairs is
swapped. By generating a permutation of the periodic arrangement in this way,
it is ensured that no node is swapped twice. The maximally disordered state is
reached after $p_{\rm max} = N/4$ swaps, which is the case used throughout this
study. There is a finite number of possible realizations, given by the
\emph{ensemble size}
\begin{equation}
    \NE = \binom{N/2}{p} \quad\rm.
    \label{eq:ensemblesize}
\end{equation}
In this study we always consider only a small subset of possible realizations,
as $\NE$ is a very large number already for the smallest considered systems.

The cumulative probability $\bp(\Nc)$ is illustrated in Fig.~\ref{fig:pm} for
various power flow thresholds $\fth = \Fth/K$ and linear system sizes $L$.  For
all considered threshold values $\fth$, the length of the tail of the
cumulative probability $\bp(\Nc)$ is increasing with increasing linear system
size $L$. Larger systems possess more consumers, so that the probability to
obtain a large number of unsupplied consumers is increasing and the values of
$\bp(\Nc)$ are increasing with system size in the tail of the distribution.

The value $\bp(0)$ marks the probability that a cascade results in one or more
unsupplied consumers. This probability is increasing for decreasing system
size and decreases with the threshold value $\fth$. This is clearly seen for low
critical values of $\fth=0.25$ and $\fth=0.3$ in Fig.~\ref{fig:pm}(a) and (b),
respectively. For $\fth = 0.4$ and $\fth = 0.5$, illustrated in
Fig.~\ref{fig:pm}(c) and (d), these trends are less clearly visible since there
are fewer observations available. Consequently, the relative standard deviation between different realizations
is also increasing with increasing threshold values.

\begin{figure}[tb]
    \includegraphics[width=\columnwidth]{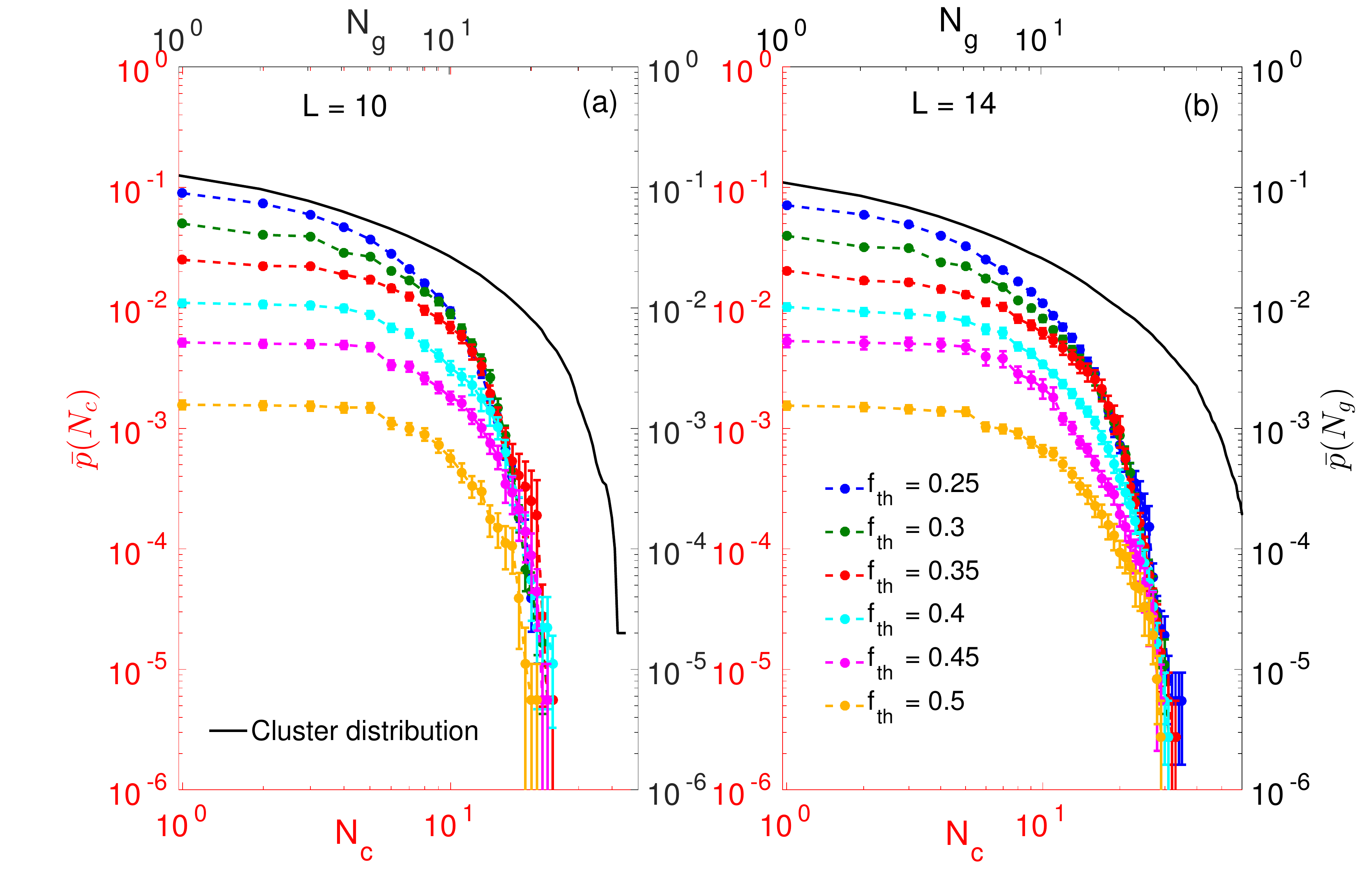}
    \caption{
        (Color online) Average cumulative probability $\bp(\Nc)$ \eqref{eq:pm}
        for square grids with varying thresholds $\fth$ and fixed linear system
        size $L$ (colored lines, red axis). The distribution of clusters of
        consumers $\bp(\Ng)$ is shown as the black line (black axis). (a) $L=10$, (b) $L=14$.
        Error bars show the standard error
        (cf.~Eq.~\eqref{eq:pm}).
    }
    \label{fig:pm-L}
\end{figure}

The results for the cumulative probabilities
$\bp(\Nc)$ in dependence of the threshold power flow $\fth$ for fixed system
size are illustrated in Fig.~\ref{fig:pm-L}. We observe an increase of
$\bp(\Nc)$ with decreasing $\fth$. The length of the tails are almost
independent of the threshold value $\fth$, indicating that even
for the largest threshold value $\fth = 0.5$ outages with a large number of unsupplied
consumers occur.

For comparison, we also identify the initially existing consumer clusters that
exist prior to the initialization of the cascade of failures. The size of an
initially present consumer cluster is denoted by $\Ng$.
The corresponding complementary cumulative probability distribution $\bp(\Ng)$
(cf.~Eqs.~\eqref{eq:er}-\eqref{eq:pm}) represents the probability that the initial
consumer clusters are larger than $\Ng$, as illustrated in Fig.~\ref{fig:pm-L}
(black curves). Interestingly, we do not find a direct relation between the
consumer cluster size of the original grid $\Ng$ and the number of unsupplied
consumers $\Nc$ after the cascade. The largest initial consumer clusters $\Ng$ are
considerably larger than the largest number of unsupplied consumers after the
cascade. For example, for the considered realizations with $L=10$, the
largest initial consumer cluster consists of 42 consumers, whereas the largest
number of unsupplied consumers after the cascade is 26. We conjecture that
in case of the square grid topology, the
distribution of originally existing consumer clusters $\Ng$ provides
an upper limit for the distribution of unsupplied consumers
$\Nc$, but we find no direct relation between $\Ng$ and $\Nc$ if the consumers
are randomly distributed.

\begin{figure}[tb]
    \includegraphics[width=\columnwidth]{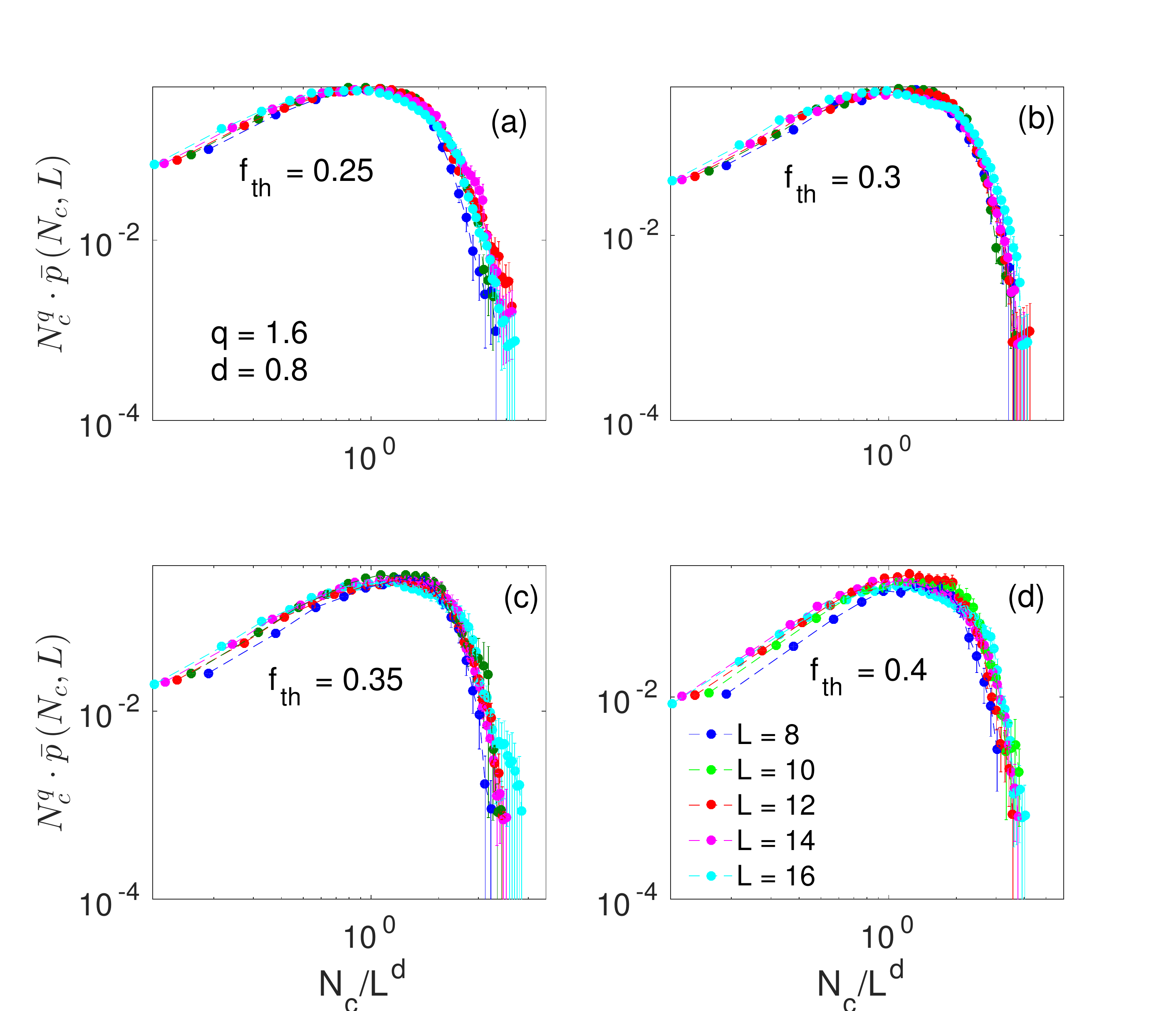}
    \caption{
        (Color online) Scaling law for $d=0.8$ and $q=1.6$. Average cumulative
        probability $\bp(\Nc, L)$ \eqref{eq:pm} for square grids of different
        linear system size $L$ and with different threshold power flow $\fth$:
        (a) $\fth = 0.25$, (b) $\fth = 0.3$, (c) $\fth = 0.35$, (d) $\fth =
        0.4$. Error bars show the standard error
        (cf.~Eq.~\eqref{eq:pm}).
    }
    \label{fig:pm-fit}
\end{figure}

\textit{Scaling Analysis.---} It has been noted in Ref.~\cite{ID} that
cascading failures may show evidence for \emph{self-organized criticality}
(SOC) \cite{bak}. SOC has been first found experimentally in rice piles, where
the distribution of the size of avalanches has been found to follow a power law
\cite{frette}. Numerical studies of avalanches in sand pile models have shown
evidence for power law behavior as well, but the accuracy of the numerical
determination of the avalanche exponents $q$ is limited by finite-size effects,
resulting in values of the order of $q=1.6$ \cite{bak,sandpile,luebeck}. The most
effective way to evaluate the numerical data is by conducting a finite-size
scaling analysis. Here we apply this strategy to evaluate our numerical
data on cascading failures, using the scaling ansatz
\begin{equation}
    \bp(\Nc,L) = \Nc^{-q} f(\Nc/L^d) \quad\rm,
    \label{eq:scaling}
\end{equation}
with some unknown scaling function $f(\Nc/L^d)$ and the effective scaling
dimension $d$.

We rescale our data according to the scaling ansatz \eqref{eq:scaling} with
parameters $d=0.8$ and $q=1.6$ illustrated in Fig.~\ref{fig:pm-fit}. For the
scaling ansatz to be valid, the curves of all system sizes $L$ should coincide.
The agreement of the data with the scaling ansatz is best for small
threshold values $\fth$. We cannot exclude that the worse agreement for
larger $\fth$ is only due to the decreasing statistics, since less cascades
are initiated.
Note that the effective scaling dimension $d=0.8$ is smaller than the spatial dimension.
We find good agreement with the scaling ansatz for values in the range $q = 1.6\pm0.2$. For
large $\fth$ we still find the best agreement with the scaling ansatz for
the same parameter values as for $\fth=0.25$. We thus find evidence for scaling of the
average cumulative probability $\bp(\Nc, L)$ with an exponent $q \approx 1.6\pm0.2$,
which possibly does not depend on the threshold value. A power law dependence
corresponds to a horizontal line of $N_c^q\bar{p}\left(N_c,L\right)$ in
Fig.~\ref{fig:pm-fit}. We only find a power law dependence for a small interval.

In contrast to the simple model of Ref.~\cite{ID} where the load $F_{ij}$ of a
failing transmission line is redistributed in equal parts among the remaining transmission
lines, we do not find evidence for a critical ratio of the threshold power $\Fth$ within the studied range of $\Fth$ values.
Interestingly, by including the physical flow model \eqref{eq:stat} in the simulation,
although we only find a small interval with a pure power law decay, the exponent $q \approx 1.6\pm 0.2$
is comparable to the one found in Ref.~\cite{ID}, $q \approx 1.4$.

\begin{figure}[t]
    \includegraphics[width=\columnwidth]{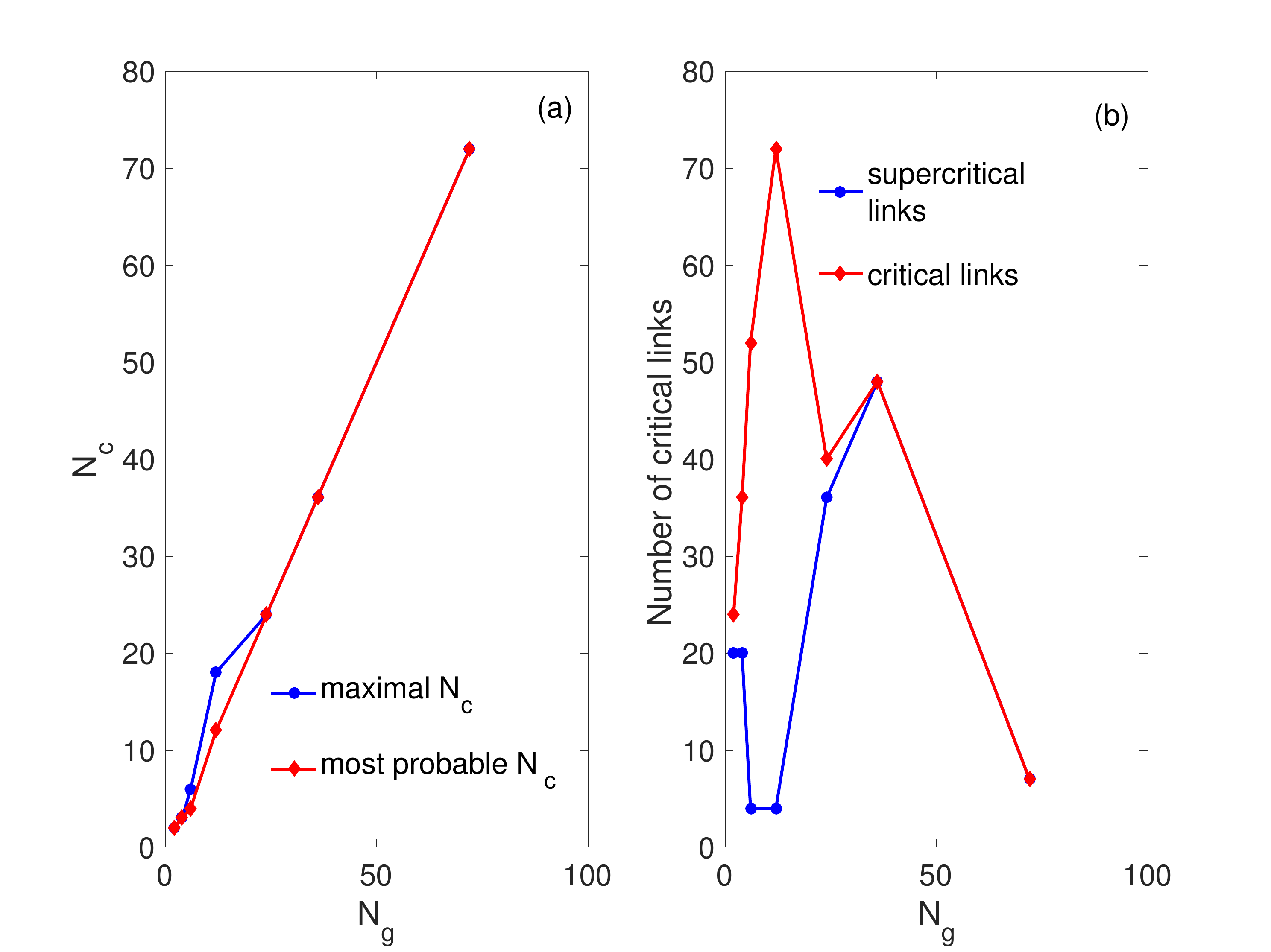}
    \vspacebeforecaption
    \caption{
        (Color online) (a) Blue: Maximal number of unsupplied consumers $\Nc$
        that occurs for all possible initial line failures as function of
        cluster size $\Ng$. Red: Most likely number of unsupplied consumers.
        (b) Blue: Number of supercritical links whose initial failure
        causes the maximal possible number of unsupplied consumers. Red:
        Number of critical links whose initial failure cause a power
        outage.
    }
    \vspaceaftercaption
    \label{fig:special1}
\end{figure}

\textit{Impact of clustering in the regular square grid topology.---} To better
understand the impact of the initial clustering of generators and consumers on
the cascading failures, we analyze regular square grids ($L=12$) with open
boundary conditions and a periodic arrangement of generator and consumer
clusters of fixed size (1x2, 2x2, 2x4, etc.). These may occur as particular
cases of the random realizations, analyzed before. For these we are able to
demonstrate a direct relation between the initial consumer cluster size $\Ng$ and
the number of unsupplied consumers $\Nc$ after the cascade. Obviously, the
largest possible clusters exist if all $N/2$ consumers are located on one side
of the grid, and all $N/2$ generators on the other side (6x12 clusters). The
smallest possible clusters are formed by pairs of connected consumers or
generators (1x2 clusters). We also consider clusters with four (2x2), eight
(2x4), twelve (3x4), 24 (4x6) and 36 (6x6) consumers or generators per cluster.
In order to obtain comparable results we determine the maximal power flow in
the initial grid $\Fmax$ and set the threshold power flow to $\Fth = \Fmax +
0.1\,{\rm s}^{-2}$.

\begin{figure*}
    \begin{center}
        \subfloat[]{\includegraphics[width=.25\textwidth]{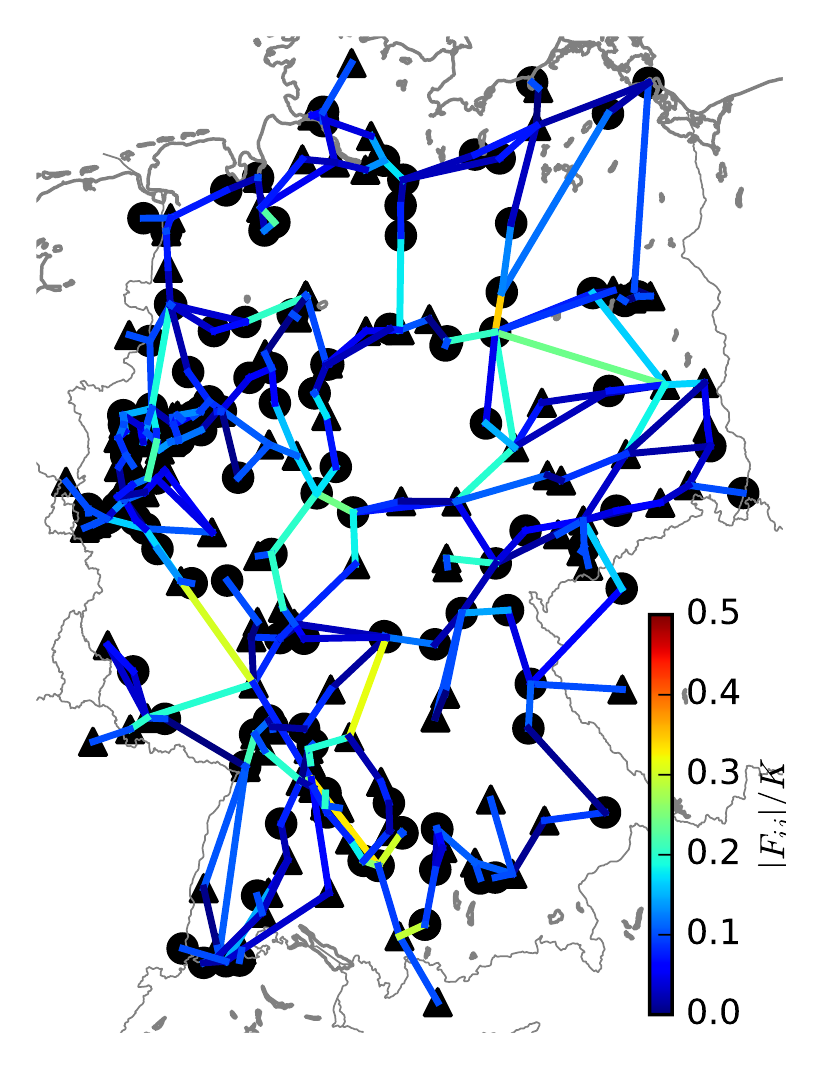}}
        \subfloat[]{\includegraphics[width=.25\textwidth]{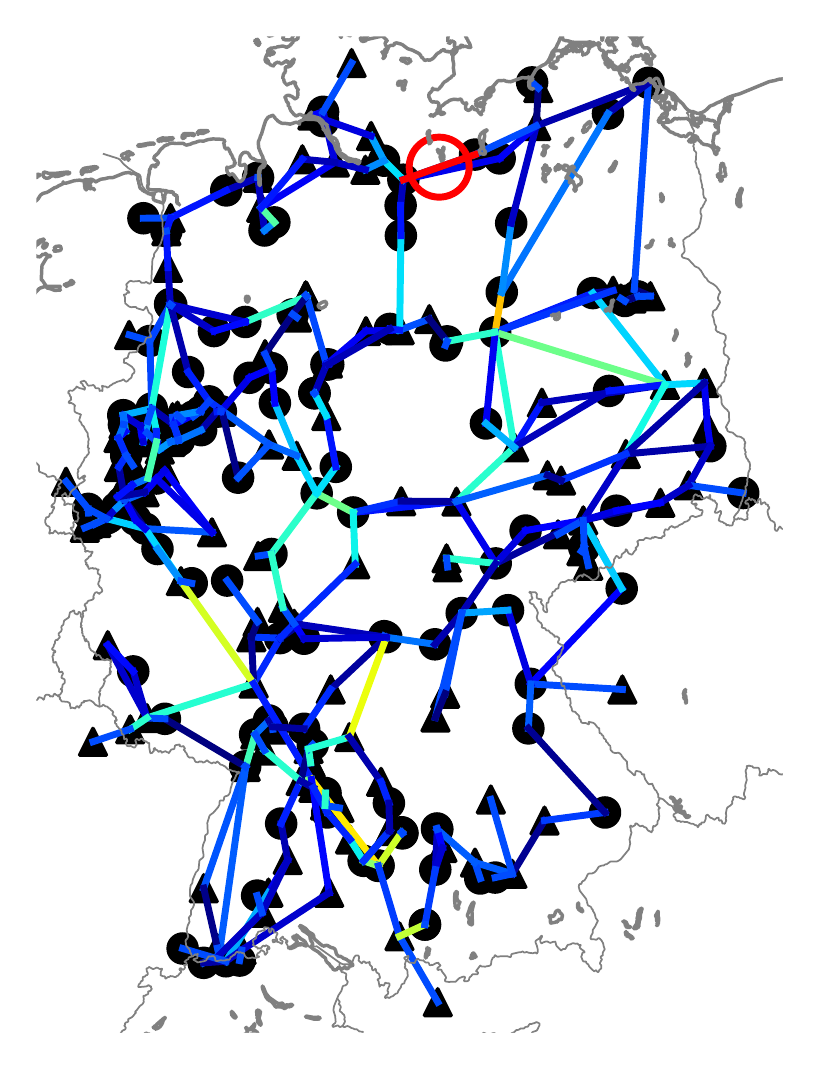}}
        \subfloat[]{\includegraphics[width=.25\textwidth]{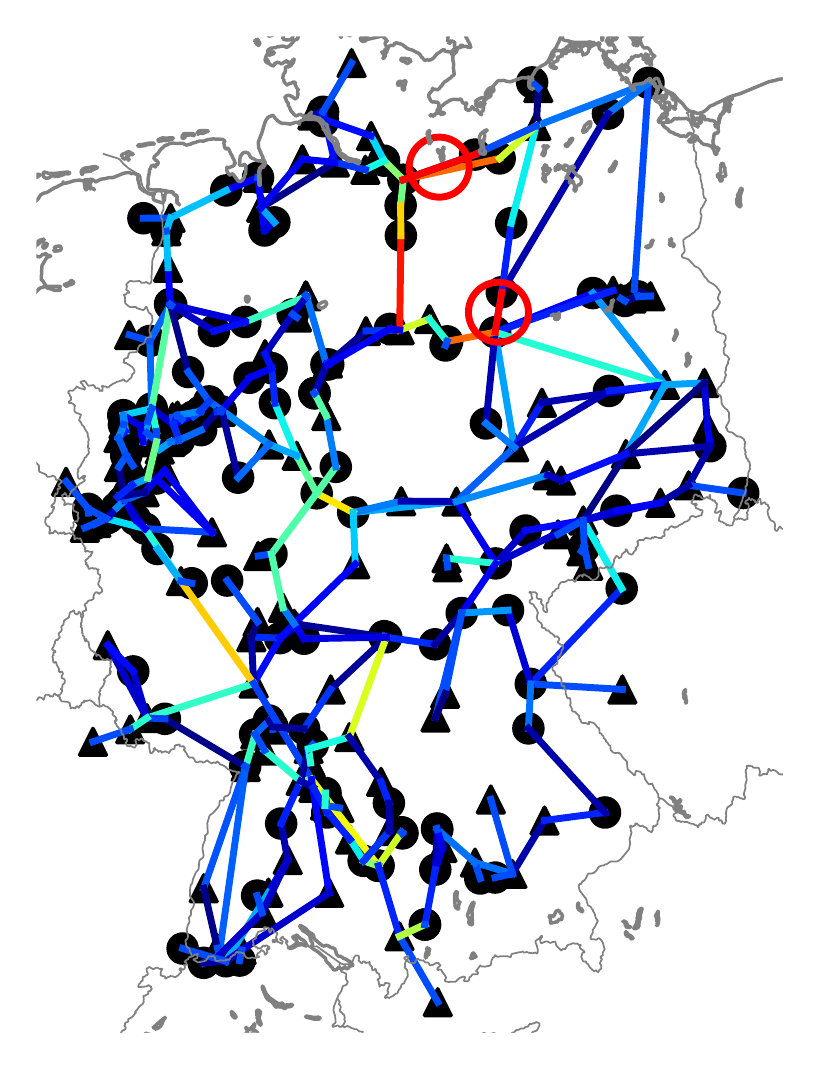}}
        \subfloat[]{\includegraphics[width=.25\textwidth]{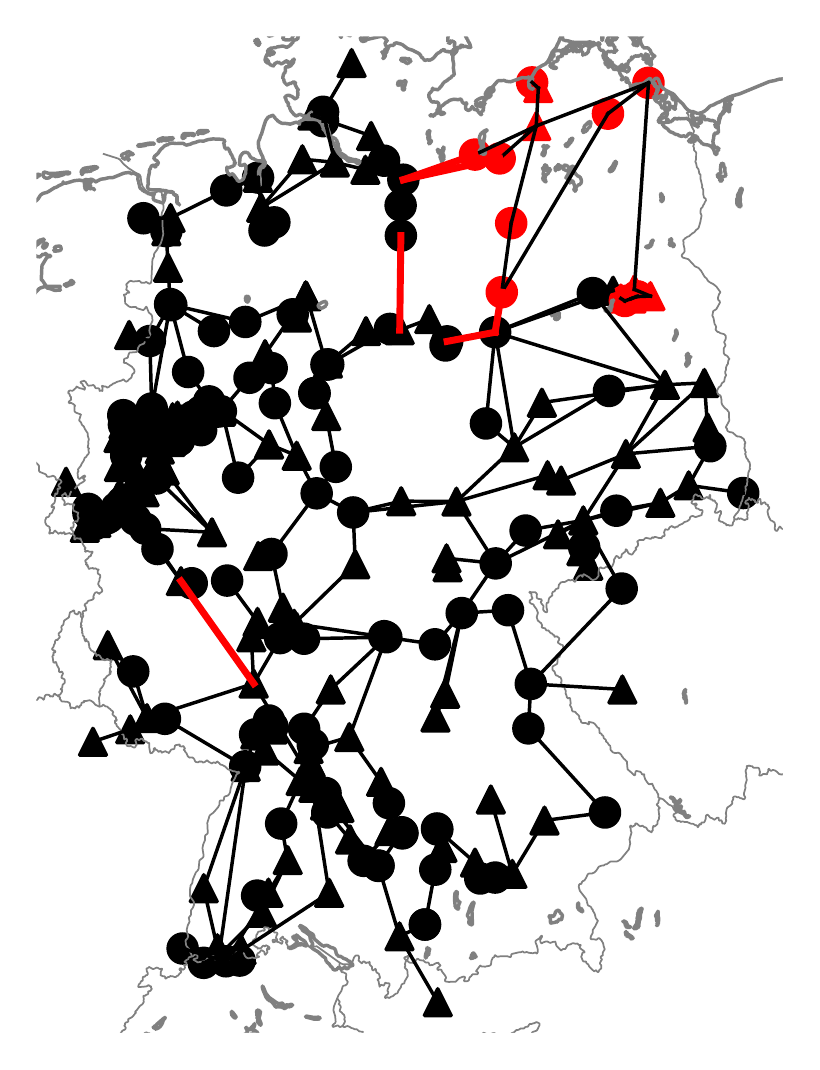}}
    \end{center}
    \vspacebeforecaption
    \caption{%
        (Color online) Example for a cascade of line failures in the German grid
        model, with {\blue transmission capacity} $K=10\,{\rm s}^{-2}$ and threshold $\fth=0.35$. Triangles denote
        generators, disks denote consumers. (a) Initial power flow. (b) The
        cascade is initiated by removing the link marked with a red circle. (c)
        Another link fails. (d) Another four links fail, and the simulation is
        stopped because the grid is not fully connected anymore. We mark the
        disconnected component containing the unsupplied nodes and other failed lines in red color.
    }
    \vspaceaftercaption
    \label{fig:cascade-germany}
\end{figure*}

We measure two quantities and study their dependence on the cluster size $\Ng$:
The maximal number of unsupplied consumers that can occur, and the most
likely number of unsupplied consumers $\Nc$ (cf.~Fig.~\ref{fig:special1}).
The results demonstrate that the most likely value of $\Nc$ increases linearly
with the cluster size $\Ng$ (cf.~Fig.~\ref{fig:special1}(a), red curve). It is
therefore the most probable event that exactly one consumer cluster disconnects
from the grid. For small cluster sizes, the maximum outage that can occur
(cf.~Fig.~\ref{fig:special1}(a), blue curve) is slightly above the most
probable outage, indicating that sometimes more than one cluster contains
unsupplied consumers at the end of the cascade. For large cluster sizes
(24..72 nodes), an outage of one cluster is at the same time the maximally
possible event and also the most probable one.

The number of critical links, i.e., those links that cause a power outage, is
found to increase with the initial cluster size $\Ng$ up to a cluster size of
eight (cf.~Fig.~\ref{fig:special1}(b), red curve). For larger cluster sizes
this number of critical links is decreasing. Links that connect different
clusters are found to be particularly vulnerable. The number of such links is
decreasing with increasing cluster size, explaining the decrease in the number
of critical links.

Next, we consider those links that cause the maximal observed outage, which we
call \emph{supercritical links}. The number of such supercritical links is
first decreasing with cluster size and than increasing again
(cf.~Fig.~\ref{fig:special1}(b), blue curve), until the number of supercritical
links eventually matches the number of critical links in the limit of large
cluster size. Only a few links cause a maximum outage for intermediate
cluster sizes, whereas for large cluster sizes almost every link causes a
maximal outage.  We conclude that large clusters are favorable in the sense
that only few transmission lines can cause an outage. On the other hand, if
such an outage occurs, the impact is much more severe, i.e., more consumers are
affected.


\begin{figure}[tb]
    \centering
      \includegraphics[width=\columnwidth]{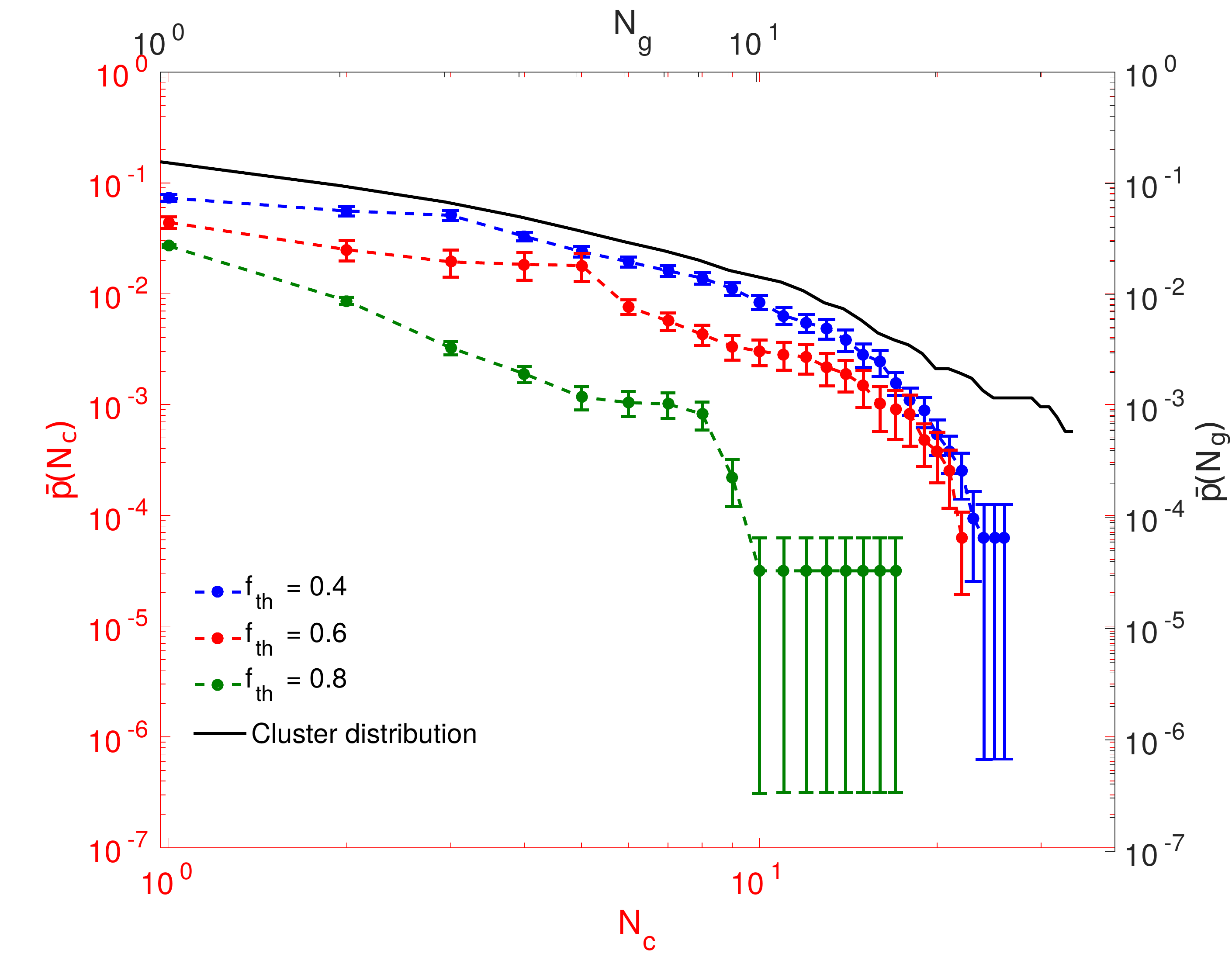}
    \caption{
        (Color online) Average cumulative probability $\bp(\Nc)$ \eqref{eq:pm}
        as function of the number of unsupplied consumers $\Nc$ for the German grid model
        for various threshold values $\fth$ (colored lines, red axis). The distribution of initial consumer clusters
        $\bp(\Ng)$ is shown as the black line (black axis). Error bars show the standard error
        (cf.~Eq.~\eqref{eq:pm}).
    }
    \label{fig:pm-ger}
\end{figure}

\textit{German Transmission Grid.---} So far, all results have been obtained
for the regular square grid topology. To test our findings on a realistic grid
topology, we consider a model for the German high-voltage transmission grid.
The model grid is based on data from the SciGRID project \cite{scigrid}, where
only the $380\,{\rm kV}$ level is considered. The objective is to test
our algorithm on a more realistic, non-regular topology, for which it is sufficient to consider
the topology of the German grid, isolated from the rest of the pan-European electricity network.
As before, we consider a binary distribution of generators and consumers and a constant line power transmission
capacity. We also apply the cascading failure algorithm described above and consider $R=1000$ realizations.
Fig.~\ref{fig:cascade-germany} demonstrates a cascade of single line failures
in the German grid model for a binary distribution of generators and consumers.
Here, we use the parameters $P = 1 \,{\rm s}^{-2}$, $K = 10 \,{\rm s}^{-2}$,
$\alpha = 1 \,{\rm s}^{-1}$, and $\fth = 0.35$.

We analyze cascading failures for various threshold values $\fth$. The results
for the cumulative probabilities $\bp(\Nc)$ are shown in Fig.~\ref{fig:pm-ger}.
With decreasing $\fth$, the values of the average cumulative probability
$\bp(\Nc)$ are increasing. This behavior is similar to that of the square grids
(cf.~Fig.~\ref{fig:pm}). Note that the German grid model contains 254 nodes and
317 links, so that its size is comparable to a $12 \times 12$ square grid,
which has 264 links, so we can compare the results with the corresponding
probability density $\bp(\Nc)$ shown in Fig.~\ref{fig:pm}. We observe clear
differences: First, for the German grid $\bp(\Nc)$ does not decay continuously
but shows abrupt steps. Second, the decay of $\bp(\Nc)$ for large $\Nc$ depends
strongly on the threshold $\fth$. For large $\fth$ the decay is clearly
exponential. For $\fth = 0.4$ the decay is slow and may indicate a power law
decay, although the statistics does not allow to extract a definite power
exponent. There might be a critical value of the threshold $\fth$ below which
the cumulative probability density becomes critical and decays with a power
law. Thus, we may conclude that the difference in the topological structure of
the German high-voltage grid as compared to the square grids leads to marked
differences in the probability of cascading failures, whose origin needs to be
studied in more detail. We also show in Fig.~\ref{fig:pm} the distribution of
consumer clusters $\bp(\Ng)$ (black line) existing in the grid before the
cascade.  We observe that for the smallest threshold value, $\fth = 0.4$, the
tail of the cumulative probability distribution of unsupplied
consumers $\Nc$ is similar to the tail of the distribution of clusters, indicating
that the probability of large outage sizes is closely related to the occurrence of
large clusters in the German grid topology.

\textit{Conclusions and Outlook.---} Single line failures can induce a cascade
of failures leading to power outages for a potentially large number of
consumers. Although cascading failures have been studied before, most of the
studies were based on simple flow models. In this article we contribute to a
more realistic model for cascading failures based on a AC load flow
model. We developed an algorithm for cascading failures which is widely
applicable to different network topologies and parameter ranges.

First, we analyzed regular square grids with a random placement of
generators and consumers. We considered different threshold values for the
transmitted power $\Fth$. We identified a scaling law with a power law exponent
for the probability to obtain a minimum number of unsupplied consumers
(blackout size) in dependence of the system size. In contrast to the simplified model of
Ref.~\cite{ID}, we do not find evidence for a critical ratio of the threshold
power $\Fth$ but rather find the same dependence in a wide range of threshold
values $\fth$. Moreover, we find a power law decay with power $q = 1.6 \pm 0.2$
which is comparable to the one found in Ref.~\cite{ID}, $q \approx 1.4$. We do
not find a direct relation between the distribution of initial clusters and the
resulting blackout sizes after the cascading failures. Often, clusters of
consumers are found to split into multiple parts during a cascade of line
failures, which might explain why the observed blackout size is often smaller
than the initial cluster size.

Second, we studied regular square grids with a periodic arrangement of consumer
clusters of fixed size. We demonstrated that large consumer clusters often lead
to large outages, whereas small clusters typically lead to only small outages
for the initial failure of a single line. In contrast, the number of critical
links, i.e., links which cause a cascade if they fail, is decreasing with
increasing cluster size. Here, we can also identify a direct relation between
consumer cluster sizes and the number of unsupplied consumers after a cascade
has been initiated.

Finally, we studied a real-world topology, a model for the German high-voltage
transmission grid. In this irregular grid structure, we find a qualitatively
similar behavior for the complementary cumulative probability $\bp(\Nc)$ that a
blackout of a certain minimum size occurs, but we also observe clear differences:
First, for the German grid, $\bp(\Nc)$ does not decay continuously but shows
abrupt steps, which could be only due to the small number of observations. Second, the decay
of $\bp(\Nc)$ for large $\Nc$ depends strongly on the threshold $\fth$. For
large $\fth$ the decay is clearly exponential. For $\fth = 0.4$ the decay is
slow and may indicate a power law decay, although the statistics does not allow
to extract a definite power exponent. There might be a critical value of
the threshold $\fth$ below which the cumulative probability density becomes
critical and decays with a power law.

Thus, we may conclude that the difference in the topological structure of the
German high-voltage grid as compared to the square grids leads to marked
differences in the probability of cascading failures. Its origin and dependence
on various topological measures will have to be studied in more detail in
future research. It remains to find criteria for the question which
arrangements of generators and consumers are beneficial for a particular grid
topology in order to minimize the chance and extent of blackouts. In future
research we will also study the influence of heterogeneous transmission line
capacities and more realistic distributions for the consumed and generated
power at each node.

\acknowledgments

We gratefully acknowledge the support of BMBF, CoNDyNet, FK.03SF0472A.


\end{document}